\documentclass{nature_figure}
\usepackage{color}
\usepackage{amsmath}
\usepackage{bm}
\usepackage{graphicx}
\title{Instability in dynamic fracture and\\ the failure of the classical theory of cracks}

\author{Chih-Hung Chen$^1$, Eran Bouchbinder$^2$ \& Alain Karma$^1$}

\begin{document}

\maketitle

\begin{affiliations}
 \item Department of Physics and Center for Interdisciplinary Research on Complex Systems, Northeastern University, Boston, Massachusetts 02115, USA
 \item Department of Chemical Physics, Weizmann Institute of Science, 7610001 Rehovot, Israel
\end{affiliations}

\begin{abstract}
Cracks, the major vehicle for material failure, tend to accelerate to high velocities in brittle materials\cite{Freund.90}. In three-dimensions, cracks generically undergo a micro-branching instability at about $40\%$ of their sonic limiting velocity\cite{Ravi-Chandar.84c, Fineberg.91, Sharon.96, Fineberg.99, livne2005universality}. Recent experiments showed that in sufficiently thin systems cracks unprecedentedly accelerate to nearly their limiting velocity without micro-branching, until they undergo an oscillatory instability\cite{Livne.07, Goldman.12}. Despite their fundamental importance and apparent similarities to other instabilities in condensed-matter physics and materials science, these dynamic fracture instabilities remain poorly understood. They are not described by the classical theory of cracks\cite{Freund.90}, which assumes that linear elasticity is valid inside a stressed material and uses an extraneous local symmetry criterion to predict crack paths\cite{Goldstein.74}. Here we develop a model of two-dimensional dynamic brittle fracture capable of predicting arbitrary paths of ultra-high-speed cracks in the presence of elastic nonlinearity without extraneous criteria. We show, by extensive computations, that cracks undergo a dynamic oscillatory instability controlled by small-scale elastic nonlinearity near the crack tip. This instability  occurs above a ultra-high critical velocity and features an intrinsic wavelength that increases proportionally to the ratio of the fracture energy to an elastic modulus, in quantitative agreement with experiments. This ratio emerges as a fundamental scaling length assumed to play no role in the classical theory of cracks, but shown here to strongly influence crack dynamics. Those results pave the way for resolving other long-standing puzzles in the failure of materials.
\end{abstract}

Crack propagation is the main mode of materials failure. It has been a topic of intense research for decades because of its enormous practical importance and fundamental theoretical interest. Despite considerable progress to date\cite{Marder.95.jmps,Adda-Bedia.99,Buehler.03,Buehler2006, Bouchbinder.10, Bouchbinder.14}, the classical theory of brittle crack propagation\cite{Freund.90} still falls short of explaining the rich dynamical behavior of high-speed cracks in brittle solids such as glass, ceramics, and other engineering, geological, and biological materials that break abruptly and catastrophically.

This theory, termed Linear Elastic Fracture Mechanics (LEFM)\cite{Freund.90}, assumes that linear elastodynamics --- a continuum version of Newton's second law together with a linear relation between stress (force) and strain (deformation) --- applies everywhere inside a stressed material except for a negligibly small region near the crack tip. It predicts the instantaneous crack velocity $v$ by equating the elastic energy release rate $G$, controlled by the intensity of the stress divergence near the crack tip, with the fracture energy $\Gamma(v)$. The scalar equation $G\!=\!\Gamma(v)$ must be supplemented with an extraneous criterion to select the crack path; the most widely used one is the Principle of Local Symmetry\cite{Goldstein.74}, which assumes that cracks propagating along arbitrary paths feature a symmetric stress distribution near their tips.

A central prediction of this theory is that straight cracks smoothly accelerate to the Rayleigh wave-speed $c_R$ (the velocity of surface acoustic waves) in large enough systems\cite{Freund.90}.
However, cracks universally undergo symmetry-breaking instabilities before reaching their theoretical limiting velocity\cite{Ravi-Chandar.84c, Fineberg.91, Sharon.96, Fineberg.99, livne2005universality, Livne.07}. In three-dimensional (3D) systems such as thick plates, instability is manifested by short-lived micro-cracks that branch out sideways from the main crack. This so-called micro-branching instability\cite{Ravi-Chandar.84c, Fineberg.91, Sharon.96, Fineberg.99, livne2005universality} typically occurs when $v$ exceeds a threshold $v_c$ of about 40\% of $c_R$. Recent experiments in brittle gels have further shown that upon reducing the thickness of the system, micro-branching is suppressed and instability is manifested at a much higher speed ($v_c\!\sim\!90\%$ of the shear wave-speed $c_s$) by oscillatory cracks with a well-defined intrinsic wavelength $\lambda$\cite{Livne.07, Goldman.12, Bouchbinder.14}. Such behavior cannot be explained by LEFM, even qualitatively, as it contains no lengthscales other than the external dimensions of the system.

To investigate dynamic fracture instabilities, we use the phase-field approach\cite{bourdin2000numerical,karma2001phase,Hakim.09,Karma_Nature.10,bourdin2014morphogenesis},
capable of describing complex crack paths while treating both short-scale material failure and large-scale elasticity, without adopting the common assumption that elasticity remains linear at arbitrary large strains near the crack tip (Fig.~1a). This generalized approach features two intrinsic length-scales missing in LEFM (Fig.~1b): the size $\xi$ of the microscopic {\it dissipation zone} around the tip, where elastic energy is dissipated while creating new fracture surfaces, and the size $\ell$ of the near-tip {\it nonlinear zone}, where linear elasticity breaks down when strains become large. We stress that while $\xi$ and $\ell$ are missing in LEFM, they are {\em consistent} with it as they remain much smaller than the size system. Moreover, $\ell$ scales with the ratio $\Gamma/\mu$ of the fracture energy to the shear modulus. Experiments and theory suggest that this nonlinear scale may be related to the oscillatory instability\cite{Livne.08,Bouchbinder.08,Bouchbinder.09a, Bouchbinder.09b,Livne.10, Goldman.12,Bouchbinder.14}, but this relationship is not fundamentally established. Here we develop a new phase-field formulation (see Methods and\cite{suppinfo}) that maintains the wave-speeds constant inside the dissipation zone, thereby avoiding spurious tip-splitting occurring in previous phase-field models at much lower crack velocities\cite{Karma.04}. This new formulation allows us to model for the first time the ultra-high-speed cracks observed experimentally.

\begin{figure}
\begin{center}
\hskip -5pt
\includegraphics[width=0.85\textwidth]{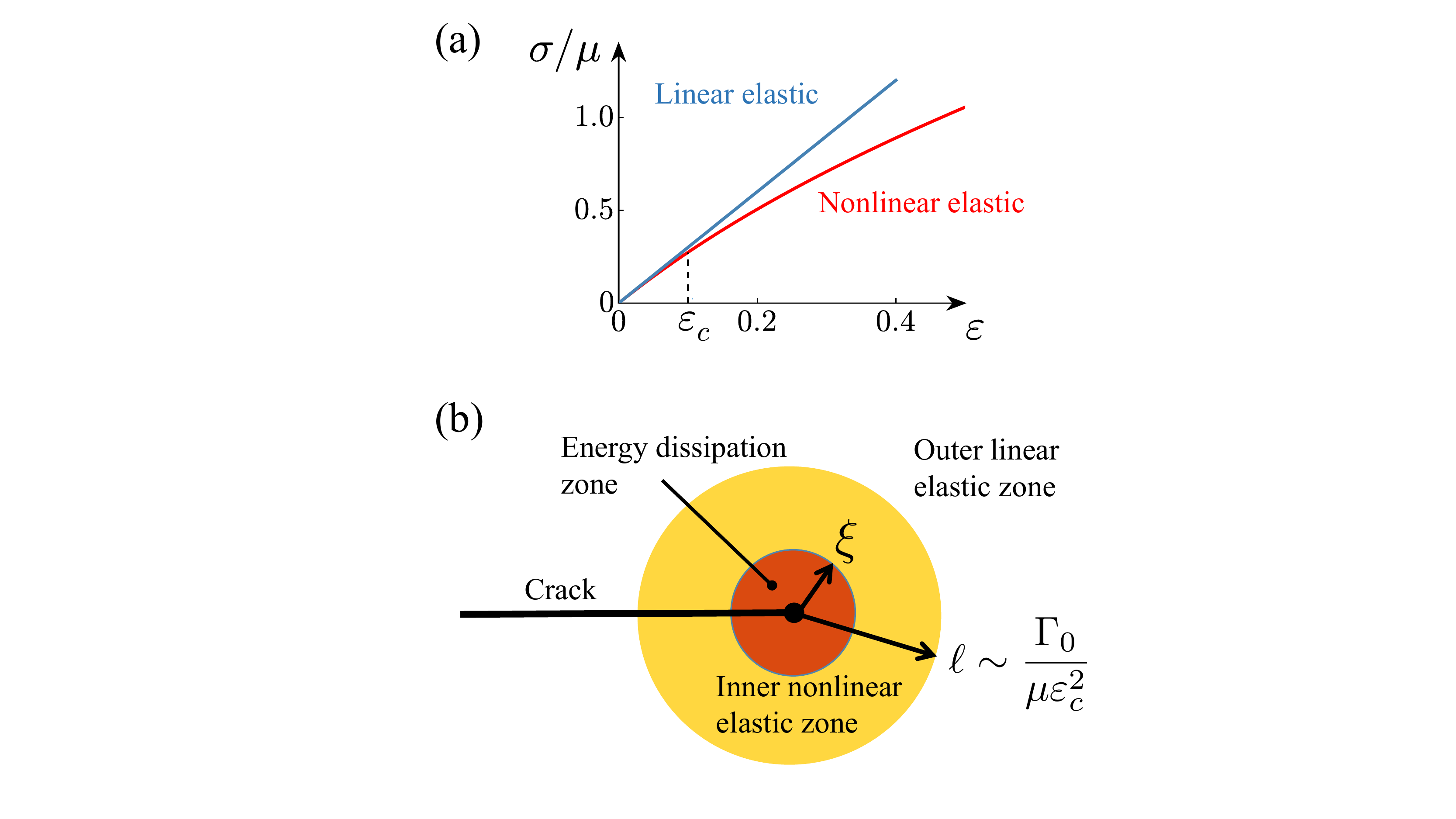}
\caption{\textbf{Nonlinear elasticity and crack tip length-scales.}
\textbf{a,} The stress $\sigma$, normalized by the shear modulus $\mu$, versus strain $\varepsilon$ for a linear elastic solid (blue line) and a nonlinear elastic solid (red line corresponding to $e_{\rm strain}$ defined by Eq. \eqref{eq:e_strain}) under uniaxial tension. Nonlinearity becomes important around $\varepsilon_c\!\approx\!0.1$.
\textbf{b,} A schematic representation of near-tip length-scales neglected in Linear Elastic Fracture Mechanics (LEFM), but consistent with it as long as they are much smaller than the system size.
These include the dissipation zone of size $\sim\!\xi$, where elastic energy is dissipated in the process of creating new crack surfaces, and the nonlinear zone of size $\sim\!\ell\!>\!\xi$, where linearity breaks down. The region where linear elasticity breaks down, termed  the  ``process zone'' in fracture mechanics, includes both the dissipation and nonlinear zones. To estimate $\ell$, note first that the LEFM stress divergence $\sigma\!\sim\!\mu\,\varepsilon\!\sim\!K_I/\sqrt{r}$, where $r$ is the distance from the crack tip and
$K_I$ is the mode-I stress intensity factor, is valid for $r$ close to, but larger than, $\ell$. As $\ell$ is the region where elastic nonlinearity becomes important,
it can be estimated by setting $r\!\sim\!\ell$ and $\varepsilon\!\sim\!\varepsilon_c$ in the last expression. Finally, invoking energy balance in the tip region\cite{Freund.90},
$\Gamma_0\!\propto\!K_I^2/\mu$, one obtains $\ell\!=\!\ell_0/\varepsilon_c^2\!\gg\!\ell_0$, with $\ell_0\!\equiv\!\Gamma_0/\mu$. 
Note that the velocity dependence of the fracture energy and the relativistic distortion of near-tip fields for crack velocities approaching $c_s$ can both have a strong influence on the size and shape of this nonlinear region, going beyond this simple estimate.}
\label{fig:Fig1}
\end{center}
\end{figure}

The nonlinear strain energy density $e_{\mbox{\scriptsize{strain}}}$ is chosen to correspond to an incompressible neo-Hookean solid\cite{suppinfo} (Fig.~1a), representing generic elastic nonlinearities and quantitatively describing the experiments of\cite{livne2005universality, Livne.07, Livne.08, Bouchbinder.14, Livne.10,Goldman.12}. We consider mode-I (tensile) cracks in strips of height $H$ (in the $y$-direction) and length $W$ (in the $x$-direction). Fixed tensile displacements $u_y(y\!=\!\pm H/2)\!=\!\pm\delta_y$ are imposed at the top and bottom boundaries with $\delta_y\!\ll\!H$ such that strains are small and linear elasticity is valid everywhere in the sample except within a small region of size $\ell\!\ll\!H$ near the crack tip, where elastic nonlinearity is important. The applied load is quantified by the stored elastic energy per unit length along $x$ in the pre-stretched intact strip, $G_0\!=\!e_{\mbox{\scriptsize{strain}}}H$, where $e_{\mbox{\scriptsize{strain}}}$ is uniquely determined by $\delta_y$.

Fig.~2 unprecedentedly demonstrates the existence of a rapid crack oscillatory instability in our simulations. Panel (a) shows a close up on the crack at the onset of oscillations in the material (undeformed) coordinates (see also Fig.~S1 and Supplementary Movie 1) and a corresponding sequence of crack snapshots in the spatial (deformed) coordinates, along with the strain energy density field. The results bear striking resemblance to the corresponding experimental observations in brittle gels\cite{Livne.07}, reproduced here in panel (b) (see also Supplementary Movie 2). Panel (c) shows the time evolution of the Cartesian components, $(v_x, v_y$), and magnitude of the crack velocity, $v\!=\!\sqrt{v_x^2+v_y^2}$, demonstrating that the instability appears when $v$ exceeds a threshold $v_c\!\approx\!0.92c_s$. Panel (d) shows the time evolution of the oscillation amplitude $A$ and wavelength $\lambda$, which both grow before saturating. The saturated amplitude is an order of magnitude smaller than the wavelength, in good agreement with experiments\cite{Livne.07}.

\begin{figure}
\begin{center}
\hskip -5pt
\includegraphics[width=0.9\textwidth]{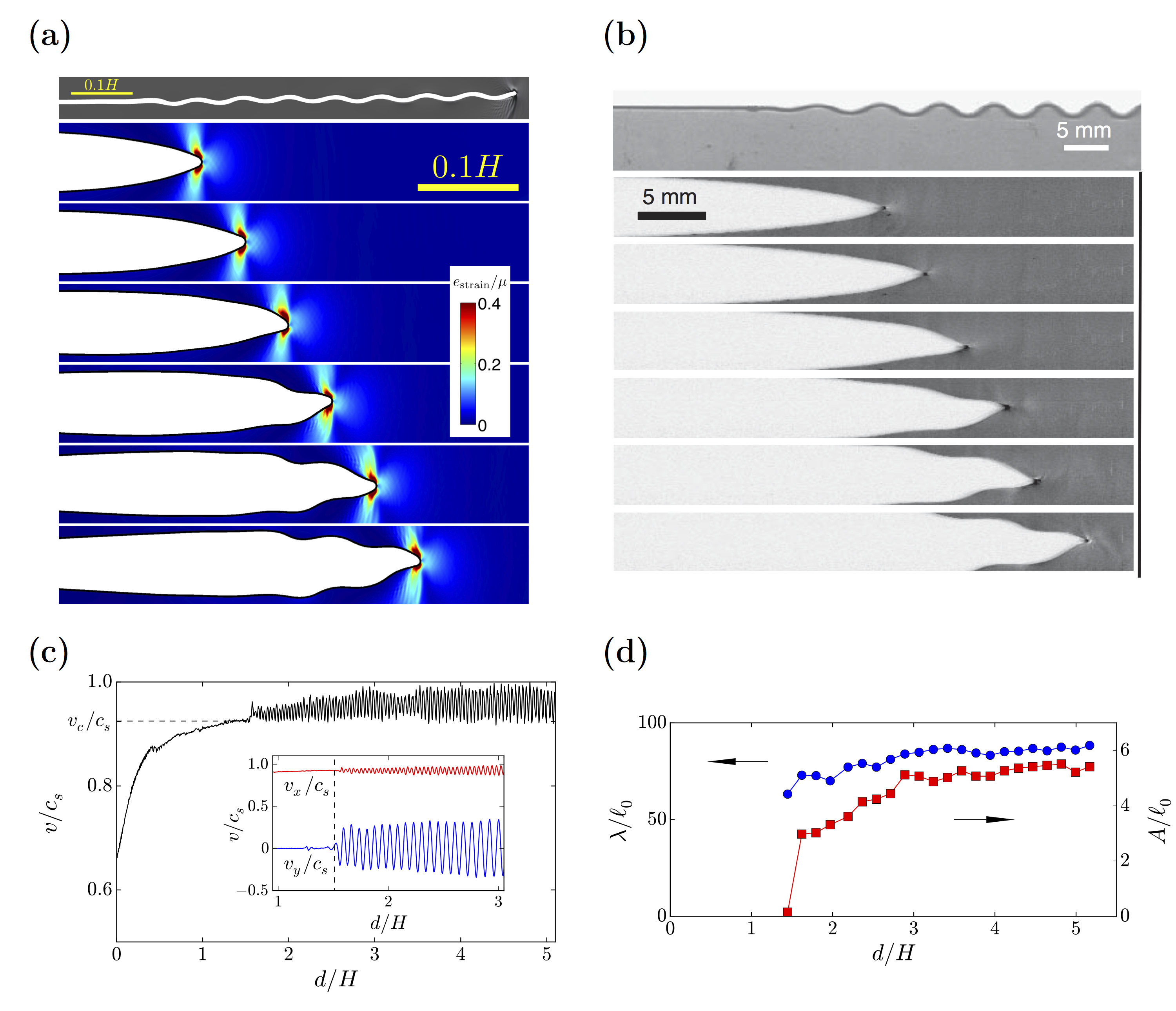}
\caption{\textbf{Onset of the oscillatory instability.}
Results of phase-field simulations of brittle mode-I (tensile) fracture in rectangular strips of height $H$ and length $W$ illustrating the onset of the oscillatory instability (full scale is available in Fig.~S1). \textbf{a,} Zoom in on the crack trajectory, defined by the $\phi\!=\!1/2$ contour,
 in the material (undeformed) frame (top) and a sequence of snapshots of the crack surfaces, along with the normalized strain energy density field
$e_{\mbox{\scriptsize{strain}}}/\mu$, in the spatial (deformed) frame (bottom). \textbf{b,} The corresponding experimental observations in brittle gels (adapted from\cite{Livne.07}). \textbf{c,} The magnitude $v$, and the cartesian components $v_x$ and $v_y$, of the crack velocity (scaled by the shear wave speed $c_s$) versus crack propagation distance $d$ (scaled by $H$). The onset of instability is marked by the dashed lines. \textbf{d,} The oscillation amplitude $A$ and wavelength $\lambda$ (scaled by $\ell_0\!=\!\Gamma_0/\mu$) versus $d/H$.
Simulation parameters are $\ell_0/\xi\!=\!0.29$, $H/\xi\!=\!300$, $W/\xi\!=\!900$, $G_0/\Gamma_0\!=\!2.6$ (corresponding to a background strain $\varepsilon_{yy}\!=\!3.62\%$), $\Delta\!=\!0.21\xi$ and $\beta\!=\!0.28$ (see Methods).}
\label{fig:Fig2}
\end{center}
\end{figure}

Importantly, we verified that the wavelength is determined by an intrinsic length-scale by carrying out simulations for different system sizes, yielding negligible variations in $\lambda$ (Fig.~S2). Moreover, we verified that the instability is caused by near-tip elastic nonlinearity by repeating the simulations using the small-strain (linear elastic) quadratic approximation of the nonlinear $e_{\mbox{\scriptsize{strain}}}$, corresponding to conventional LEFM. The latter yielded straight cracks that tip-split upon surpassing a velocity of $\approx\!0.9c_s$, without oscillations. Since both forms of $e_{\mbox{\scriptsize{strain}}}$ --- nonlinear neo-Hookean and its small-strain linear elastic approximation --- are nearly identical everywhere in the system outside the near-tip nonlinear zone, we conclude that nonlinearity within this zone is at the heart of the oscillatory instability.

To investigate the dependence of the oscillatory instability on the external load and material properties, we varied $G_0/\Gamma_0$, where $\Gamma_0\!\equiv\!\Gamma(v\!=\!0)$ is the fracture energy at onset of crack propagation, and the material-dependent ratio $\Gamma_0/(\mu\xi)\!\equiv\!\ell_0/\xi$ controls the relative strength of near-tip elastic nonlinearity and dissipation. Results of extensive simulations are shown in Fig.~3, presenting the crack velocity versus propagation distance for several $G_0/\Gamma_0$ values, and $\ell_0/\xi\!=\!0.29$ in Fig. 3a (also used in Fig. 2) and $\ell_0/\xi\!=\!1.45$ in Fig. 3b. The plots clearly show that the onset of instability occurs when $v$ exceeds a threshold value $v_c$ independently of the external load. Decreasing the load simply reduces the crack acceleration and hence $v$ exceeds $v_c$ after a larger propagation distance; instability is not observed for the lowest loads in panels (a) and (b) because $v$ has not yet reached $v_c$ by the end of the simulations. Comparing panels (a) and (b), $v_c$ is seen to increase by only a few percent when $\Gamma_0/(\mu\xi)$ is increased fivefold. This result is consistent with the experimental finding that $v_c$ remains nearly constant when the ratio of the fracture energy to the shear modulus is varied several fold by altering the composition of the brittle gels, which affects both quantities and hence their ratio\cite{Livne.07}. Furthermore, the onset velocity $v_c\!\approx\!0.92\,c_s$ in Figs.~2 and 3a is in
remarkably good quantitative agreement with experiments\cite{Livne.07}.

\begin{figure}
\begin{center}
\hskip -5pt
\includegraphics[width=0.9\textwidth]{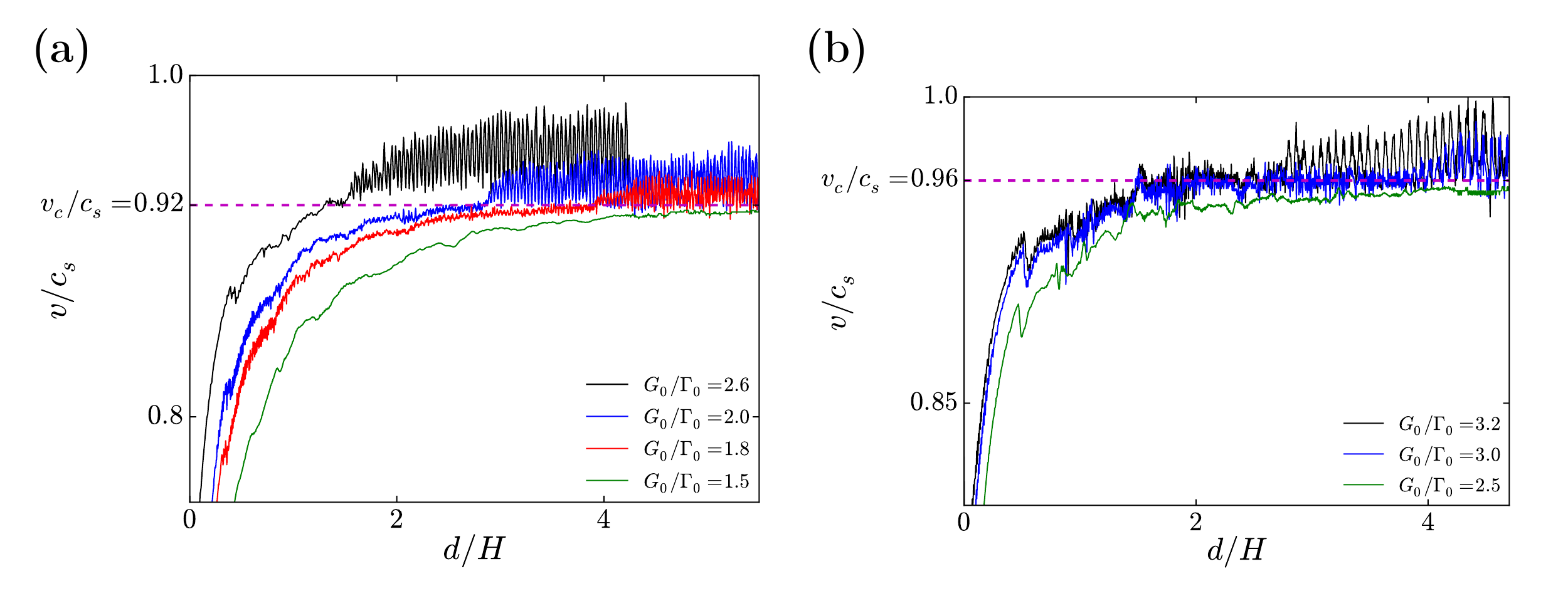}
\caption{\textbf{Critical velocity of instability.} Results demonstrating the independence of the critical velocity of instability $v_c$ on the external loading and the ratio of fracture energy to the shear modulus, which controls the size of the near-tip nonlinear zone. The normalized crack velocity $v/c_s$ versus the normalized crack propagation distance $d/H$ for different external loads, measured by the dimensionless ratio $G_0/\Gamma_0$, and
$\ell_0/\xi\!=\!0.29$ (in \textbf{a}) and $\ell_0/\xi\!=\!1.45$ (in \textbf{b}). The critical velocity in \textbf{b} exceeds slightly $c_R=0.933\, c_s$ due to elastic nonlinearity\cite{suppinfo}.
Other simulations parameters are $H/\xi\!=\!300$, $W/\xi\!=\!300$ (calculations with $W/\xi\!=\!900$ yielded the same critical velocity $v_c$), $\Delta\!=\!0.21\xi$ and $\beta\!=\!0.28$.}
\label{fig:Fig3}
\end{center}
\end{figure}

Unlike the critical velocity of instability $v_c$, the wavelength of oscillations $\lambda$ has been experimentally observed to significantly vary when the fracture energy $\Gamma(v_c)$ and the shear modulus $\mu$ were changed by varying the material composition\cite{Goldman.12}. The phase-field framework allows to independently vary $\Gamma(v_c)$ and $\mu$, and also to assess the role of the energy dissipation scale $\xi$. The size of the nonlinear zone $\ell$ is theoretically expected to be proportional to (and much larger than) $\Gamma(v_c)/\mu$, but as the pre-factor is not sharply defined, we plot in Fig.~4a $\lambda$ versus $\Gamma(v_c)/\mu$, both scaled by $\xi$, where $\Gamma(v_c)/\mu\approx 1.2\, \Gamma_0/\mu$ is obtained from accelerating crack simulations (see Methods).
We superimpose in Fig.~4a experimental measurements in brittle gels, where a value $\xi\!\simeq\!153\,\mu$m was chosen to match the $y$-intercepts of linear best fits of both the theoretical and experimental results. The slopes $d\lambda/d(\Gamma(v_c)/\mu)$, which are independent of the choice of $\xi$, are in remarkably good quantitative agreement. This agreement demonstrates that small-scale elastic nonlinearity, which is quantitatively captured by the phase-field approach, is a major determinant of the oscillation wavelength $\lambda$ that increases linearly with $\Gamma(v_c)/\mu$, when $\Gamma(v_c)/(\mu\,\xi)$ is sufficiently large. The existence of finite $y$-intercepts further suggests that the dynamics on the dissipation scale also affects $\lambda$. All in all, these results demonstrate the failure of the classical theory of fracture, which assumes that these intrinsic length-scales play no role in crack propagation under strongly dynamic conditions.

\begin{figure}
\begin{center}
\hskip -10pt
\includegraphics[width=0.8\textwidth]{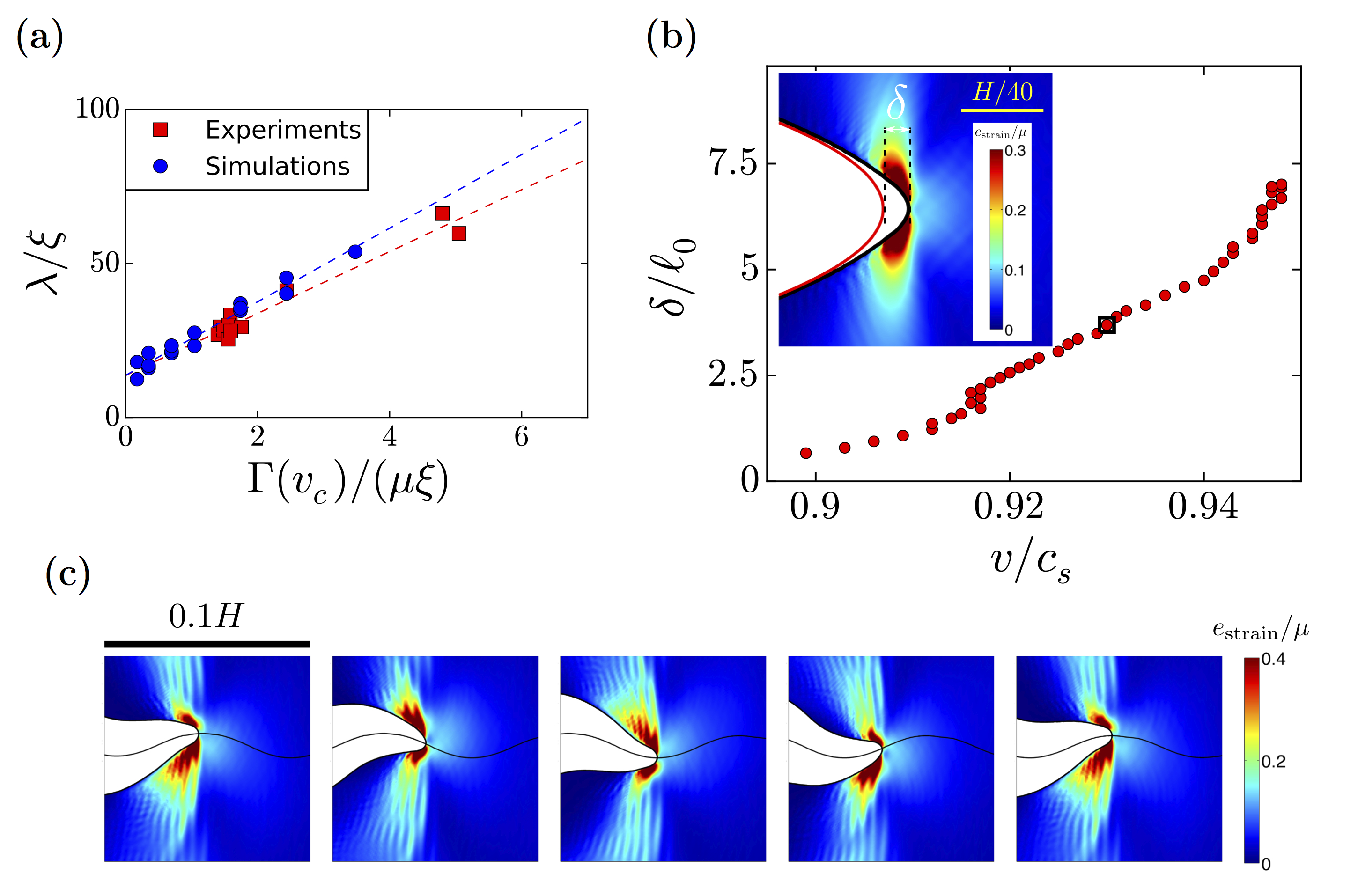}
\caption{\textbf{Critical wavelength of instability and crack tip dynamics.} Results demonstrating a strong dependence of the wavelength of instability $\lambda$
 on the ratio of the fracture energy $\Gamma(v_c)$ to the shear modulus $\mu$, which controls the size of the near-tip nonlinear zone. \textbf{a,} The wavelength $\lambda$ versus  $\Gamma(v_c)/\mu$ (both quantities scaled by $\xi$). Experimental measurements in brittle gels (data extracted from\cite{Goldman.12}) are superimposed on this plot by setting $\xi\!\simeq\!153\,\mu$m such that the $y$-intercepts of the best linear fits (dashed lines) of the theoretical and experimental results coincide. This $\xi$ value is roughly consistent with experimental estimates\cite{Livne.10}.
Note that the theoretical data  span a wider range of values compared to the experimental data, and do so in a much more continuous manner. The reason for this is that in the
experiments $\Gamma(v_c)$, $\mu$ and $\xi$ are varied in a correlated (and not well-controlled) manner through varying the material composition, while in the theory these
physical quantities can be varied independently, capable of representing a broad range of different materials.
\textbf{b,} (main) The normalized crack tip nonlinear deviation $\delta/\ell_0$ from the parabolic LEFM tip (see inset for a visual definition) versus the normalized crack velocity $v/c_s$
 and (inset) a snapshot of the crack surfaces and normalized strain energy density $e_{\mbox{\scriptsize{strain}}}/\mu$ in the spatial (deformed) coordinates for $v/c_s\!=\!0.93$ (marked by the black square in the main panel). \textbf{c,} A sequence of snapshots of the crack surfaces and the normalized strain energy density $e_{\mbox{\scriptsize{strain}}}/\mu$
during one complete steady-state oscillation cycle, demonstrating that the asymmetry in the near-tip strain fields is temporally out-of-phase with the instantaneous crack propagation direction (see also Supplementary Movie 1). Simulation parameters in \textbf{a} are the same as in Fig.~3 except the ratio $\ell_0/\xi$ varying between 0.15 and  2.9. Simulation parameters in \textbf{b} are $\ell_0/\xi\!=\!1.45$, $H/\xi\!=\!300$, $W/\xi\!=\!900$, $G_0/\Gamma_0\!=\!3.0$ (corresponding to a background strain $\varepsilon_{yy}\!=\!8.90\%$), $\Delta = 0.21\xi$ and $\beta\!=\! 0.28$. Simulation parameters in \textbf{c} are the same as in Fig.~2.}
\label{fig:Fig4}
\end{center}
\end{figure}

Elastic nonlinearity has been found experimentally to also affect the crack tip shape\cite{Livne.08,Bouchbinder.09a,Livne.10}, which strongly departs at high velocities from the parabolic shape predicted by LEFM. This departure is quantified by the deviation $\delta$ of the actual tip location from its predicted location based on the parabolic shape. Fig.~4b shows that $\delta$ indeed grows dramatically in a narrow range of ultra-high velocities approaching $v_c$ and can be larger than $\ell_0$ in agreement with experiments\cite{Livne.08,Livne.10,Goldman.12}. During oscillations, the crack-tip shape and near-tip nonlinear elastic fields become asymmetrical about the instantaneous crack propagation axis on a scale comparable to $\delta$, as illustrated in Fig.~4c, which shows snapshots of the crack tip shape and strain energy density during one complete steady-state oscillation cycle. This illustrative sequence reveals that the asymmetry in the near-tip strain fields is temporally out-of-phase with the instantaneous crack propagation direction, signaling a breakdown of the Principle of Local Symmetry under dynamic conditions\cite{Bouchbinder.09b}. How asymmetry on the scale of the nonlinear zone provides an instability mechanism, e.g.~the one proposed in\cite{Bouchbinder.09b}, remains to be further elucidated. Our newly developed nonlinear phase-field model of high-speed cracks provides a unique framework to address this and other fundamental issues, such as the basic relationship of crack oscillations and 3D micro-branching suggested by recent experiments\cite{Goldman.15}.

\begin{methods}

We use the phase-field framework\cite{bourdin2000numerical,karma2001phase} that couples the evolution of the material displacement vector field ${\bm u}(x,y,t)$ to a scalar field $\phi(x,y,t)$ which varies smoothly in space between the fully broken ($\phi\!=\!0$) and pristine ($\phi\!=\!1$) states of the material. The present model is formulated in terms of the Lagrangian $L\!=\!T\!-\!U$ where
\begin{eqnarray}
U = \int \left[\frac{\kappa}{2} \left(\nabla\phi\right)^{2}+ g(\phi)\left(e_{\mbox{\scriptsize{strain}}} - e_{c}\right)\right]dV \qquad\hbox{and}\qquad T= \frac{\rho}{2}\displaystyle \int  f(\phi)  \frac{\partial {\bm u}}{\partial t}\cdot  \frac{\partial {\bm u}}{\partial t} dV \label{kinetic}
\end{eqnarray}
represent the potential and kinetic energy, respectively, $\rho$ is the mass density inside the pristine material, and $dV$ is a volume element.
The form of $U$ implies that the broken state ($\phi\!=\!0$) becomes energetically favored when the strain energy density $e_{\mbox{\scriptsize{strain}}}$ exceeds a threshold $e_c$, and the function $g\left(\phi\right)\!=\!4\phi^{3}-3\phi^{4}$ is a monotonously increasing function of $\phi$ that controls the softening of elastic energy at large strains. The parameters $\kappa$ and $e_c$ determine the size of the dissipation zone $\xi\!=\!\sqrt{\kappa/(2e_{c})}$ and the fracture energy $\Gamma_0\!=\!2\sqrt{2\kappa e_{c}}\int_{0}^{1}d\phi\sqrt{1-g(\phi)}\!\approx\!2.9e_c\xi$ in the quasi-static limit\cite{karma2001phase,Hakim.09}. The evolution equations for $\phi$ and ${\bm u}$ are derived from Lagrange's equations
\begin{eqnarray}
\frac{\partial}{\partial t}\left[\frac{\delta L}{\delta\left(\partial\psi/\partial t\right)}\right]-\frac{\delta L}{\delta\psi}
+\frac{\delta D}{\delta\left(\partial\psi/\partial t\right)}=0, \label{Lagrange}
\end{eqnarray}
for $\psi=(\phi,u_x,u_y)$, where the functional
\begin{equation}
D=\frac{1}{2\chi}\int \left(\frac{\partial\phi}{\partial t}\right)^{2}dV \label{dissipation}
\end{equation}
controls the rate of energy dissipation. As shown in\cite{suppinfo}, it follows from Eqs.~\eqref{kinetic}-\eqref{dissipation} that $d(T+U)/dt\!=\!-2D\!\le\!0$. 
This gradient flow condition implies that the total energy (kinetic + potential) decreases monotonously in time due to energy dissipation near the crack tip where $\phi$ varies. In addition, we impose the standard irreversibility condition $\partial_t\phi\!\le\!0$.

The above model distinguishes itself from previous phase-field models\cite{Karma.04,Henry.08} by the formulation of the kinetic energy in Eq.~\eqref{kinetic}. Previous models exhibit a tip-splitting instability in a velocity range (40\% to 55\% of $c_s$ depending on the mode of fracture and Poisson's ratio\cite{Karma.04,Henry.08}) much lower than the velocity in which the oscillatory instability is experimentally observed\cite{Livne.07}. Furthermore, tip-splitting generates two symmetrically branched cracks which are qualitatively distinct from both the 3D micro-branching and 2D oscillatory instabilities. Tip-splitting can be suppressed by choosing $f(\phi)$ in Eq.~\eqref{kinetic} to be a monotonously increasing function of $\phi$, similarly to $g(\phi)$. In particular, $f(\phi)\!=\!g(\phi)$ (used here) ensures that the wave-speeds remain constant inside the dissipation zone, which is physically consistent with the fact that dissipation and structural changes near crack tips in real materials
do not involve large modifications of the wave-speeds. As the latter control the rate of transport of energy in the dissipation zone, cracks in this model can accelerate to unprecedented velocities approaching $c_s$, as observed experimentally in quasi-2D geometries\cite{Livne.07}.

In addition, unlike conventional phase-field models, we focus on a nonlinear strain energy density given by 
\begin{eqnarray}
e_{\mbox{\scriptsize{strain}}}=\frac{\mu}{2}\,\left(F_{ij}F_{ij}+[\det({\bm F})]^{-2}-3\right), \label{eq:e_strain}
\end{eqnarray}
where $F_{ij}\!=\!\delta_{ij}+\partial_{j}u_{i}$ are the components of the deformation gradient tensor and $i,j\!=\!\{x,y\}$.
It corresponds to a 2D incompressible neo-Hookean constitutive law,
exhibiting generic elastic nonlinearities and quantitatively describing the brittle gels in experiments\cite{livne2005universality,  Livne.10}.
In the small strain limit, neo-Hookean elasticity reduces to standard linear elasticity with a shear modulus $\mu$ and a 2D Poisson's ratio $\nu\!=\!1/3$.

The equations are nondimensionalized by measuring length in units of $\xi$ and time in units of $\tau\!=\!1/\left(2\chi e_c\right)$, characterizing the timescale of energy dissipation. Crack dynamics is then controlled
by only two dimensionless parameters: $e_c/\mu$ and $\beta\!\equiv\!\tau c_s/\xi$. The first controls the ratio $\ell_0/\xi=2.9e_c/\mu$, where $\ell_0\!=\!\Gamma_0/\mu$ sets the size of the nonlinear zone where elasticity breaks down (Fig.~1). The second controls the velocity dependence of the fracture energy. In the ideal brittle limit, $\beta\!\ll\!1$, $\Gamma(v)\!\approx\!\Gamma_0$ is independent of $v$. In the opposite limit, $\beta\!\gg\!1$, dissipation is sluggish and $\Gamma(v)$ is a strongly
increasing function of $v$. We vary $e_c/\mu$ between $0.05$ and $1.0$ to control the importance of elastic nonlinearity and choose a value $\beta\!=\!0.28$ so that $\Gamma(v)/\Gamma_0$ increases by about $20\%$ when $v$ varies from zero to $v_c$ (see the inset of Fig. S4), similarly to experiments\cite{Livne.10}. The equations are discretized in space on a uniform square mesh with a grid spacing $\Delta\!=\!0.21\xi$ and finite-difference approximations of spatial derivatives, and integrated in time using a Beeman's scheme\cite{suppinfo} with a time-step size $\Delta t\!=\!5\!\times\!10^{-4} \tau$. Large scale simulations of $10^6\!-\!10^7$ grid points are performed using graphics processing units (GPUs) with the CUDA parallel programming language.

LEFM has been experimentally validated for accelerating cracks that follow straight trajectories,
prior to the onset of instabilities\cite{sharon1999confirming, Bouchbinder.14}.
Therefore, as a quantitative test of our high-speed crack model\cite{suppinfo}, we verified the predictions of LEFM by showing that an accelerating crack centered inside a strip (as illustrated in Fig.~S3) satisfies the scalar equation of motion $G\!=\!\Gamma(v)$, as long as their trajectory remains straight ($v\!<\!v_c$). 
We performed this test for the nonlinear form of $e_{\mbox{\scriptsize{strain}}}$ defined by Eq. \eqref{eq:e_strain} by monitoring the instantaneous total crack length $a$ (distinct from the crack tip propagation distance $d$ plotted in Figs. 2 and 3), crack velocity $v$, and energy release rate $G$ calculated directly by contour integration using the $J$-integral\cite{suppinfo}. The results show that cracks accelerated under different loads $G_0/\Gamma_0$ exhibit dramatically different $v$ versus $a$ curves (Fig. S4), but the same $G$ versus $v$ curves (inset of Fig. S4) that defines a unique fracture energy $\Gamma(v)$ (independent of the external load and crack acceleration history). Those results are consistent with the theoretical expectation that the relation $G\!=\!\Gamma(v)$, which simply accounts for energy balance near the crack tip, should remain valid even in the presence of elastic nonlinearity as long as $G$ is calculated through the $J$-integral evaluated in a non-dissipative region.

To investigate crack instabilities we carried out simulations using a treadmill method\cite{Karma_Nature.10,Karma.04} that maintains the crack tip in the center of the strip by periodically adding a strained layer at the right vertical boundary ahead of the crack tip and removing a layer at the opposite left boundary. This method allows us to study large crack propagation distances by mimicking an infinite strip with negligible influence of boundary effects.

\end{methods}


\bibliography{fracture_Nature}



\begin{addendum}
 \item[Code availability]
The CUDA implementation of the phase-field simulation code is available upon request.
 \item[Acknowledgements] This research was supported by the U.S.-Israel Binational Science Foundation (BSF), grant no.~2012061. E.B. acknowledges support from the William Z. and Eda
Bess Novick Young Scientist Fund and the Harold Perlman Family. The authors thank Matteo Nicoli for his contribution to the initial development of the phase-field simulation code.
 \item[Competing Interests]  
 The authors declare that they have no
competing financial interests.
\item[Author contributions]  
All authors contributed equally to this work.
 \item[Correspondence]  
Correspondence and requests for materials
should be addressed to Alain Karma.~(email: a.karma@neu.edu).
\end{addendum}


\end{document}


\maketitle
\begin{affiliations}
 \item Department of Physics and Center for Interdisciplinary Research on Complex Systems, Northeastern University, Boston, Massachusetts 02115, USA
 \item Department of Chemical Physics, Weizmann Institute of Science, 7610001 Rehovot, Israel
\end{affiliations}

{\bf How ultra-high-speed cracks propagate in materials that break suddenly ?}

Brittle materials such as glass, ceramic plates, or bone break suddenly by the propagation of fast cracks that concentrate stresses at their tips. While cracks are theoretically predicted to accelerate along straight trajectories to a very high velocity close to the sound speed, they are universally observed to become unstable before reaching this speed and follow complex paths for reasons that are not fundamentally understood. This work uses state-of-the-art computational models of brittle fracture to reveal the root cause of instability in a setting where modeling predictions can be quantitatively compared to experiments for the first time. 

{\bf Movies of oscillatory crack instability in dynamic fracture ---}
The accompanying movies reveal the onset of oscillatory instability of ultra-fast tensile cracks propagating in phase-field simulations of brittle fracture in striking agreement with experimental observations. 
The Supplementary Movie 1 (simulation\_oscillations.mp4) shows a propagating tensile crack obtained by numerical simulation of the phase-field model described in this supplementary material, undergoing a transition from a straight to an oscillatory trajectory upon surpassing a critical velocity of about 92\% of the shear wave-speed, in quantitative agreement with the experiments.
The Supplementary Movie 2 (experiment\_oscillations.mp4, courtesy of Ariel Livne, Tamar Goldman and Jay Fineberg, Racah Institute of Physics, Hebrew University of Jerusalem, Israel) shows a tensile crack propagating in a brittle polyacrylamide gel and undergoing a transition from a straight to an oscillatory trajectory upon surpassing a critical velocity of about 90\% of the shear wave-speed. The movie was generated by illuminating the sample from the back and imaging the propagating crack with a high speed camera. Details about the material and the experiments can be found in\cite{Livne.07}. 

{\bf Phase-field model ---} As explained in the {\em Methods} section of the manuscript, the phase-field framework\cite{bourdin2000numerical,karma2001phase} couples the evolution of the vector displacement field ${\bm u}(x,y,t)$ to a scalar field $\phi(x,y,t)$, which varies smoothly in space between constant values corresponding to the fully broken ($\phi\!=\!0$) and pristine ($\phi\!=\!1$) states of the material. The present model, as detailed in the {\em Methods} section, is formulated in terms of the Lagrangian $L\!=\!T-U$, where
\begin{eqnarray}
U = \int \left[\frac{\kappa}{2}\left(\nabla\phi\right)^{2}+ g(\phi)\left(e_{\mbox{\scriptsize{strain}}} -e_{c}\right)\right]dV \label{potential}
\end{eqnarray}
and
\begin{eqnarray}
T= \frac{\rho}{2}\displaystyle \int  f(\phi)   \frac{\partial {\bm u}}{\partial t}\cdot  \frac{\partial {\bm u}}{\partial t} dV \label{kinetic}
\end{eqnarray}
represent the potential and kinetic energy, respectively, $\rho$ is the mass density inside the pristine material, and $dV$ is a volume element. The functions $g(\phi)$ and $f(\phi)$ will be discussed below.

The form of $U$ implies that the broken state ($\phi\!=\!0$) becomes energetically favored when the strain energy density $e_{\mbox{\scriptsize{strain}}}$ exceeds a threshold $e_c$.  The function $g(\phi)\!=\!4\phi^{3}-3\phi^{4}$ is a monotonously increasing function of $\phi$ that controls the softening of the elastic moduli at large strains such that $g(0)\!=\!0$, $g(1)\!=\!1$ and $g'(1)\!=\!g'(0)\!=0$\cite{karma2001phase, hakim2009laws}. The parameters $\kappa$ and $e_c$ determine the size of the dissipation zone $\xi\!=\!\sqrt{\kappa/\left(2e_{c}\right)}$ and the fracture energy $\Gamma_0\!=\!2\sqrt{2\kappa e_{c}}\int_{0}^{1}d\phi\sqrt{1-g\left(\phi\right)}\!\approx\! 2.9e_c\xi$ in the quasi-static limit\cite{karma2001phase,hakim2009laws}. Energy dissipation takes place on a characteristic time scale $\tau\!=\!1/(2\chi e_c)$ (see Eq.~\eqref{dissipation} for the introduction of $\chi$) in the dissipation zone of size $\xi$ around the crack tip, where $\phi$ varies smoothly between $0$ and $1$\cite{Karma.04}. The iso-surfaces of $\phi\!=\!1/2$ conventionally define the crack surfaces.

The evolution equations for $\phi$ and $\bm u$ are derived from Lagrange's equations
\begin{eqnarray}
\frac{\partial}{\partial t}\left[\frac{\delta L}{\delta\left(\partial\psi/\partial t\right)}\right]-\frac{\delta L}{\delta\psi}
+\frac{\delta D}{\delta\left(\partial\psi/\partial t\right)}=0, \label{Lagrange}
\end{eqnarray}
for $\psi=(\phi,u_x,u_y)$, where the functional
\begin{equation}
D=\frac{1}{2\chi}\int \left(\frac{\partial\phi}{\partial t}\right)^{2}dV\label{dissipation}
\end{equation}
controls the rate of energy dissipation.

Previous 2D phase-field models that adopted the above formulation with $f(\phi)\!=\!1$, gave rise to tip-splitting at marginally dynamic crack propagation velocities of $0.40\!-\!0.55c_s$ (depending on the mode of fracture and Poisson's ratio)\cite{Karma.04,Henry.08}. These results appear to be inconsistent with experiments which either reveal an intrinsically 3D micro-branching instability at this range of velocities\cite{Sharon.96, Fineberg.99, livne2005universality, Bouchbinder.14,Ravi-Chandar.84c} or cracks that accelerate to much higher velocities and undergo an oscillatory instability in the quasi-2D limit of thin samples\cite{Livne.07, Bouchbinder.14, Goldman.12}.
In 2D phase-field models that use $f(\phi)\!=\!1$, the wave speed within the dissipation zone drops to zero --- because the elastic moduli degrade and eventually vanish as $\phi$ drops to zero, while the mass density remains fixed --- thus limiting the rate of energy transport to the crack tip. However, realistic dissipation processes --- such as plastic deformation near crack tips --- are not accompanied by significantly reduced wave speeds in this region.

To address this serious concern, we adopt the above formulated model with $f(\phi)$ that degrades with $\phi$ similarly to $g(\phi)$, disallowing large variations of the wave speeds in the dissipation zone compared to the surrounding elastic medium. In particular, we use $f(\phi)\!=\!g(\phi)$ for simplicity, ensuring that the wave speed inside the dissipation zone remains constant. This feature, which distinguishes the present model from previous ones, is shown to suppress the spurious tip-splitting instability and to allow cracks to accelerate to unprecedentedly high velocities approaching $c_s$, as is observed experimentally in quasi-2D geometries\cite{Livne.07}.

{\bf Coupled dissipative and elastodynamics ---}  Using Eqs.~\eqref{potential}-\eqref{dissipation}, one can derive the following equations of motion for the phase and displacements fields
\begin{eqnarray}
\frac{1}{\chi}\frac{\partial\phi}{\partial t}  &=& -\frac{\delta U}{\delta\phi}+\frac{\rho}{2}\,\frac{\partial f}{\partial\phi} \frac{\partial {\bm u}}{\partial t}\cdot  \frac{\partial {\bm u}}{\partial t},\label{eq:eom-phi}\\
\rho\,f\,\frac{\partial^{2}u_x}{\partial t^{2}}&=&-\frac{\delta U}{\delta u_x}-\rho\,\frac{\partial f}{\partial t}\frac{\partial u_x}{\partial t},\\
\rho\,f\,\frac{\partial^{2}u_y}{\partial t^{2}}&=&-\frac{\delta U}{\delta u_y}-\rho\,\frac{\partial f}{\partial t}\frac{\partial u_y}{\partial t}.
 \label{eq:eom-u}
\end{eqnarray}
In addition, the time derivatives of the potential and kinetic energy (cf.~Eqs.~\eqref{potential}-\eqref{kinetic}) take the form
\begin{eqnarray}
\frac{dU}{dt}  = \int\left(\frac{\delta U}{\delta\phi}\frac{\partial\phi}{\partial t}+\sum_{i=x,y}\frac{\delta U}{\delta u_{i}}\frac{\partial u_{i}}{\partial t}\right)dV  \label{eq:dUdt}
\end{eqnarray}
and 
\begin{eqnarray}
\frac{dT}{dt}=\rho \int\left[\frac{1}{2}\frac{\partial f}{\partial t} \frac{\partial {\bm u}}{\partial t}\cdot  \frac{\partial {\bm u}}{\partial t}
+f \frac{\partial {\bm u}}{\partial t}\cdot  \frac{\partial^2 {\bm u}}{\partial t^2}\right] dV, \label{eq:dTdt}
 \end{eqnarray}
respectively.

Substituting expressions for $\delta U/\delta\phi$ and $\delta U/\delta u_i$ obtained from Eqs.~\eqref{eq:eom-phi}-\eqref{eq:eom-u} into Eq.~\eqref{eq:dUdt}, one obtains
\begin{eqnarray}
\frac{dU}{dt} & = & \int\left(\frac{\delta U}{\delta\phi}\frac{\partial\phi}{\partial t}+\sum_{i=x,y}\frac{\delta U}{\delta u_{i}}\frac{\partial u_{i}}{\partial t}\right)dV\nonumber \\
 & = & -\frac{1}{\chi}\int\left(\frac{\partial\phi}{\partial t}\right)^{2}dV+\frac{\rho}{2}\,\int\frac{\partial f}{\partial t} \frac{\partial {\bm u}}{\partial t}\cdot  \frac{\partial {\bm u}}{\partial t}dV -\rho\int\left(f\frac{\partial^{2}{\bm u}}{\partial t^{2}}
 +\frac{\partial f}{\partial t}\frac{\partial {\bm u}}{\partial t}\right)\cdot\frac{\partial {\bm u}}{\partial t}dV\nonumber \\
 & = & -\frac{1}{\chi}\int\left(\frac{\partial\phi}{\partial t}\right)^{2}dV-\frac{dT}{dt}, \label{eq:budget}
\end{eqnarray}
where Eq.~\eqref{eq:dTdt} was used in the last step. Hence, we have
\begin{eqnarray}
\frac{d}{dt}\left(T + U\right) = -\frac{1}{\chi}\int\left(\frac{\partial\phi}{\partial t}\right)^{2}dV =-2D \le 0,\label{dissipative}
\end{eqnarray}
demonstrating that the system follows gradient flow dynamics in which the total energy (kinetic + potential) decreases monotonously in time due to energy dissipation near the crack tip where $\phi$ varies. In addition, we impose the standard irreversibility condition $\partial_t\phi\!\le\!0$, which prevents cracks from healing.

{\bf Nonlinear elasticity and application to neo-Hookean solids ---} Another distinguishing feature of the present approach is that we allow the strain energy density $e_{\mbox{\scriptsize{strain}}}$ to be nonlinear. The physical rationale, as argued in\cite{livne2008breakdown, bouchbinder2008weakly, Bouchbinder.09a, Livne.10}, is that dissipation (irreversible deformation) is generically expected to be preceded by nonlinear elastic (reversible) deformation. That is, with decreasing distance from the crack tip, linear elasticity first gives way to nonlinear elasticity and then to dissipation in the formation of new fracture surfaces. To make a specific choice of elastic nonlinearity, we focus on a 2D plane-stress incompressible neo-Hookean energy functional
\begin{eqnarray}
e_{\mbox{\scriptsize{strain}}}=\frac{\mu}{2}\,\left(F_{ij}F_{ij}+[\det({\bm F})]^{-2}-3\right), \label{eq:e_strain}
\end{eqnarray}
where $F_{ij}\!=\!\delta_{ij}+\partial_{j}u_{i}$ are the components of the deformation gradient tensor ($i,j\!=\!\{x,y\}$) and $[\det({\bm F})]^{-1}$ is the out-of-plane stretch ratio.
It is motivated by the fact that it is one of the simplest nonlinear generalizations of linear elasticity and because it quantitatively describes the thin brittle polyacrylamide gels used in the experiments of\cite{livne2005universality, livne2008breakdown, Livne.10}, the same materials in which the oscillatory instability of ultra-fast cracks has been observed\cite{Livne.07, Goldman.12}.

The stress tensor that is thermodynamically conjugate to $\bm F$ (the so-called first Piola-Kirchhoff stress tensor\cite{Holzapfel}) is given by ${\bm s}\!=\!\partial_{{\bm F}}\,e_{\mbox{\scriptsize{strain}}}$. Using it, the $J$-integral\cite{Freund.90, Nakamura1985, Bouchbinder.09a, Livne.10} --- which quantifies the instantaneous rate of energy flowing through any closed contour $C$ surrounding the crack tip --- is given as
\begin{eqnarray}
{\cal J}=\int_{C}\left[\left(U+T\right)v\,n_{x}+s_{ij}\frac{\partial u_{i}}{\partial t}n_{j}\right]dC, \label{J-integral}
\end{eqnarray}
where $ \bm n$ is the outward unit vector normal to $C$ and $v$ is crack velocity. The energy release rate $G$ is then given by
\begin{eqnarray}
G = {\cal J}/v. \label{J-integral2}
\end{eqnarray}
Finally, note that in the small strain limit, neo-Hookean elasticity reduces to standard linear elasticity with a shear modulus $\mu$ and a 2D Poisson's ratio $\nu\!=\!1/3$\cite{Bouchbinder.2010_IJF}.

{\bf Large-scale simulations ---} We nondimensionalize the model equations by measuring length in units of the dissipation zone size $\xi$ and time in units of the characteristic energy dissipation time $\tau\!=\!1/(2\chi e_c)$. In those units, crack dynamics is controlled by only two dimensionless parameters: $e_c/\mu$ and $\beta\!\equiv\!\tau c_s/\xi$. The scalar phase-field $\phi$ and vector displacement field ${\bm u}$ are discretized in space on a uniform square mesh with a grid spacing $\Delta\!=\!0.21\xi$ and finite-difference approximations of spatial derivatives, and integrated in time using a Beeman's scheme\cite{beeman1976} with a time-step size of $\Delta t/\tau\!=\!5\times10^{-4}$.  $e_c/\mu$ is varied between $0.05$ and $1.0$ to control the importance of elastic nonlinearity. Large scale simulations of the order of $10^6\!-\!10^7$ grid points are performed using graphics processing units (GPUs) with the CUDA parallel programming language.

{\bf Strip geometry and loading conditions ---} The present phase-field approach was first used to investigate the stability of mode-I (tensile) cracks in rectangular strips of height $H$ (in the $y$-direction) and length $W$ (in the $x$-direction). A seed crack of length $W/2$ propagating along the $x$-axis, with its tip at the center of the strip, is initially inserted. Tensile loading is imposed by fixed displacements $u_y(y\!=\!\pm H/2)\!=\!\pm\delta_y$ at the top and bottom boundaries with $\delta_y\!\ll\!H$ such that strains remain small (from $2\!-\!3\%$ up to $15\%$). These small loads ensure that the linear approximation to $e_{\mbox{\scriptsize{strain}}}$ is valid everywhere in the sample except within a small region of size $\ell\!\ll\!H$ near the crack-tip, where elastic nonlinearity is important. The applied load is quantified by the total stored elastic energy in the pre-stretched intact strip per unit length along $x$, $G_0\!=\!e_{\mbox{\scriptsize{strain}}}H$, where $e_{\mbox{\scriptsize{strain}}}$ is uniquely determined by $\delta_y$.

We employ a treadmill method\cite{Karma_Nature.10,Karma.04} to maintain the crack tip in the center of the strip by periodically adding a strained layer at the right vertical boundary ahead of the crack tip and removing a layer at the opposite left boundary. This method allows to simulate crack tip propagation distances $d$ much larger than the strip length $W$ to mimic infinite strip conditions with negligible influence of boundary effects. Fig.~S1 shows a dynamic crack undergoing an oscillatory instability when its velocity $v$ exceeds a threshold $v_c$, as discussed extensively in the manuscript.

\begin{figure}
\begin{center}
\hskip -5pt
\includegraphics[width=0.8\textwidth]{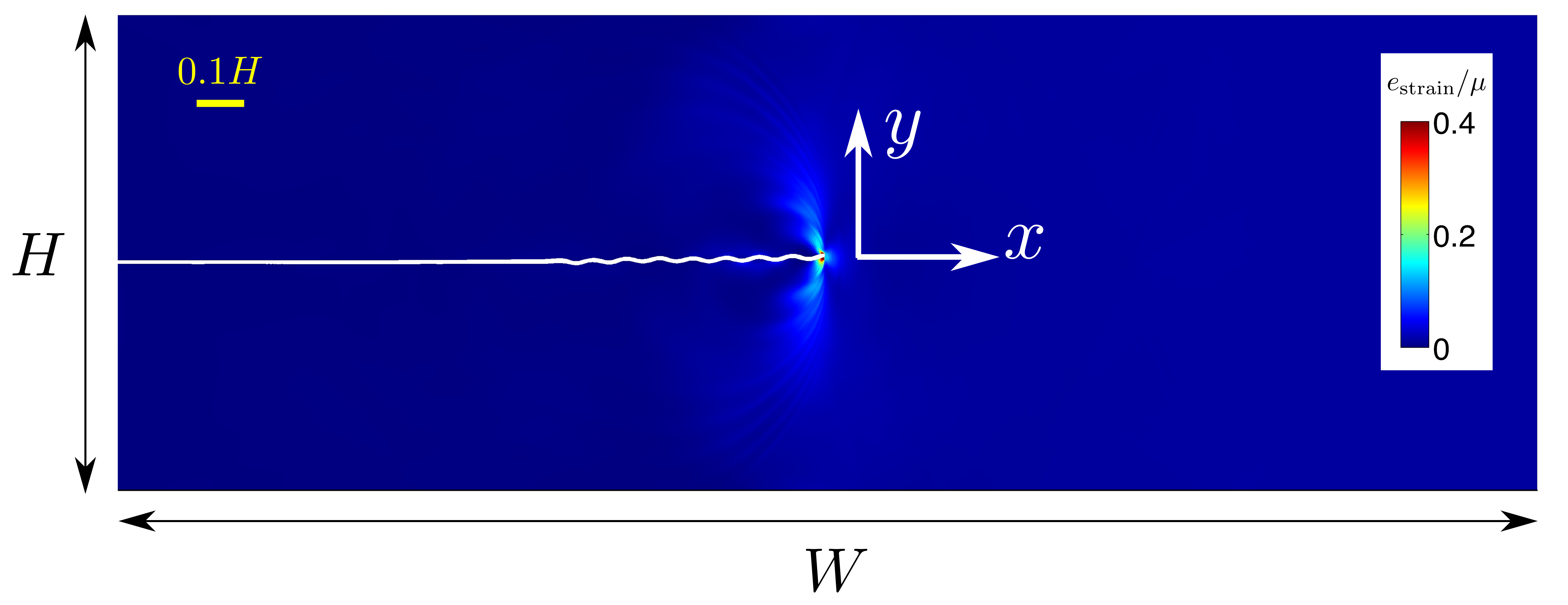}
\caption{{\bf Onset of the oscillatory instability.} A snapshot a large-scale phase-field simulation of dynamic crack propagation in a rectangular strip of height $H$ (in the $y$-direction) and length $W$ (in the $x$-direction), directly demonstrating a transition from a straight path to an oscillatory one when the crack velocity $v$ exceeds a critical velocity $v_c\!\approx\!0.92 c_s$. Simulation parameters are $\ell_0/\xi\!=\!0.29$, $H/\xi\!=\!300$, $W/\xi\!=\!900$, $\beta\!=\!0.28$, $\Delta\!=\!0.21\xi$ and $G_0/\Gamma_0\!=\!2.6$ corresponding to a background strain $\varepsilon_{yy}\!=\!3.62\%$. A treadmill method is used, as explained in the text.}
\label{figs1}
\end{center}
\end{figure}
The system height $H$ is systematically varied in order to verify that the wavelength of oscillations is an intrinsic length-scale independent of extrinsic/geometric length-scales. Indeed, Fig.~S2 clearly demonstrates that the variations in the wavelength are negligible when $H$ is varied fourfold.
\begin{figure}
\begin{center}
\hskip -5pt
\includegraphics[width=0.7\textwidth]{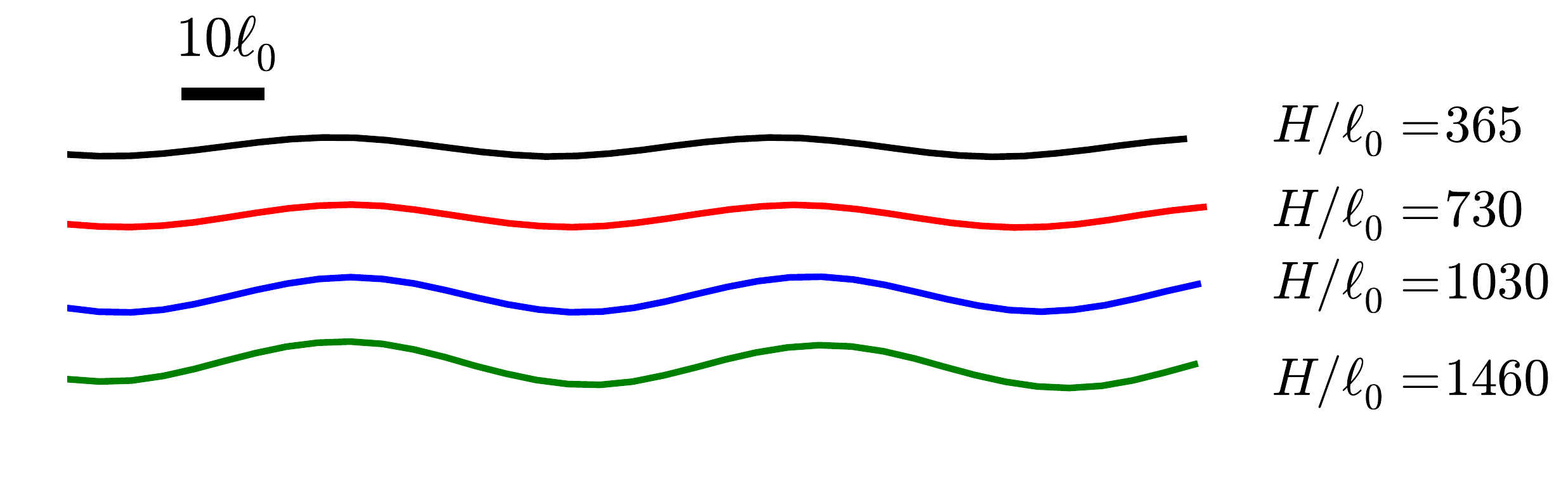}
\caption{
{\bf System size independence.} A few oscillatory crack paths obtained from phase-field simulations with different system heights $H/\ell_0\!=\!365, 730, 1030, 1460$. The results clearly demonstrate that the wavelength of oscillations exhibits negligible variation with $H$, which is varied fourfold. This shows that the wavelength of oscillations is an intrinsic length-scale, independent of extrinsic/geometric length-scales. Simulation parameters are $G_0/\Gamma_0\!=\!1.8$, $\ell_0/\xi\!=\!0.29$, $\beta\!=\!0.28$, $\Delta\!=\!0.21\xi$ and $W\!=\!1030\ell_0$.
}
\label{figs2}
\end{center}
\end{figure}


{\bf Verification of the equation of motion for straight cracks from classical LEFM theory ---} The scalar equation of motion $G\!=\!\Gamma(v)$ from classical LEFM theory should be valid as long as the propagating crack remains straight. This prediction can serve as a test case for the validity of our proposed framework. To that aim, we use the nonlinear $e_{\mbox{\scriptsize{strain}}}$ defined by Eq.~\eqref{eq:e_strain} and carry out crack propagation simulations for various loading levels $G_0/\Gamma_0\!=\!2, 3, 4, 6$ in the strip geometry of Fig.~S3.

\begin{figure}
\begin{center}
\hskip -5pt
\includegraphics[width=1.04\textwidth]{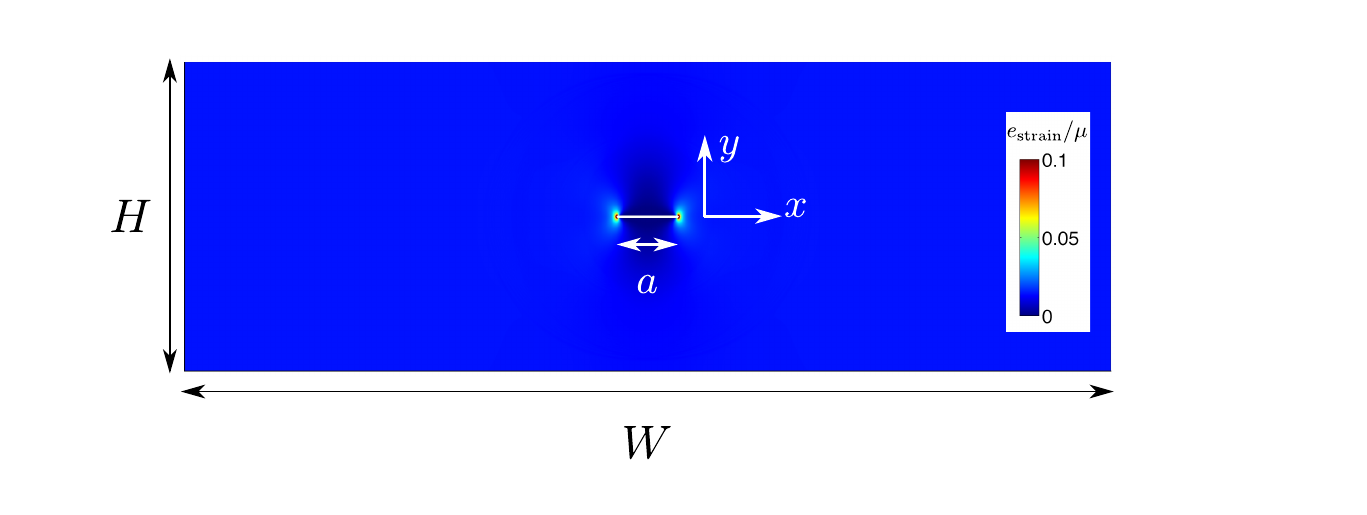}
\caption{{\bf Strip geometry for accelerating crack simulations.}
Snapshot of phase-field simulation showing a crack of instantaneous length $a$ propagating in a strip of height $H$ (in the $y$-direction) and length $W$ (in the $x$-direction) along the symmetry line in the $x$-direction. The strip is loaded by fixed displacements $u_y(y\!=\!\pm H/2)\!=\!\pm\delta_y$.}
\label{figs3}
\end{center}
\end{figure}

\begin{figure}
\begin{center}
\hskip -5pt
\includegraphics[width=0.65\textwidth]{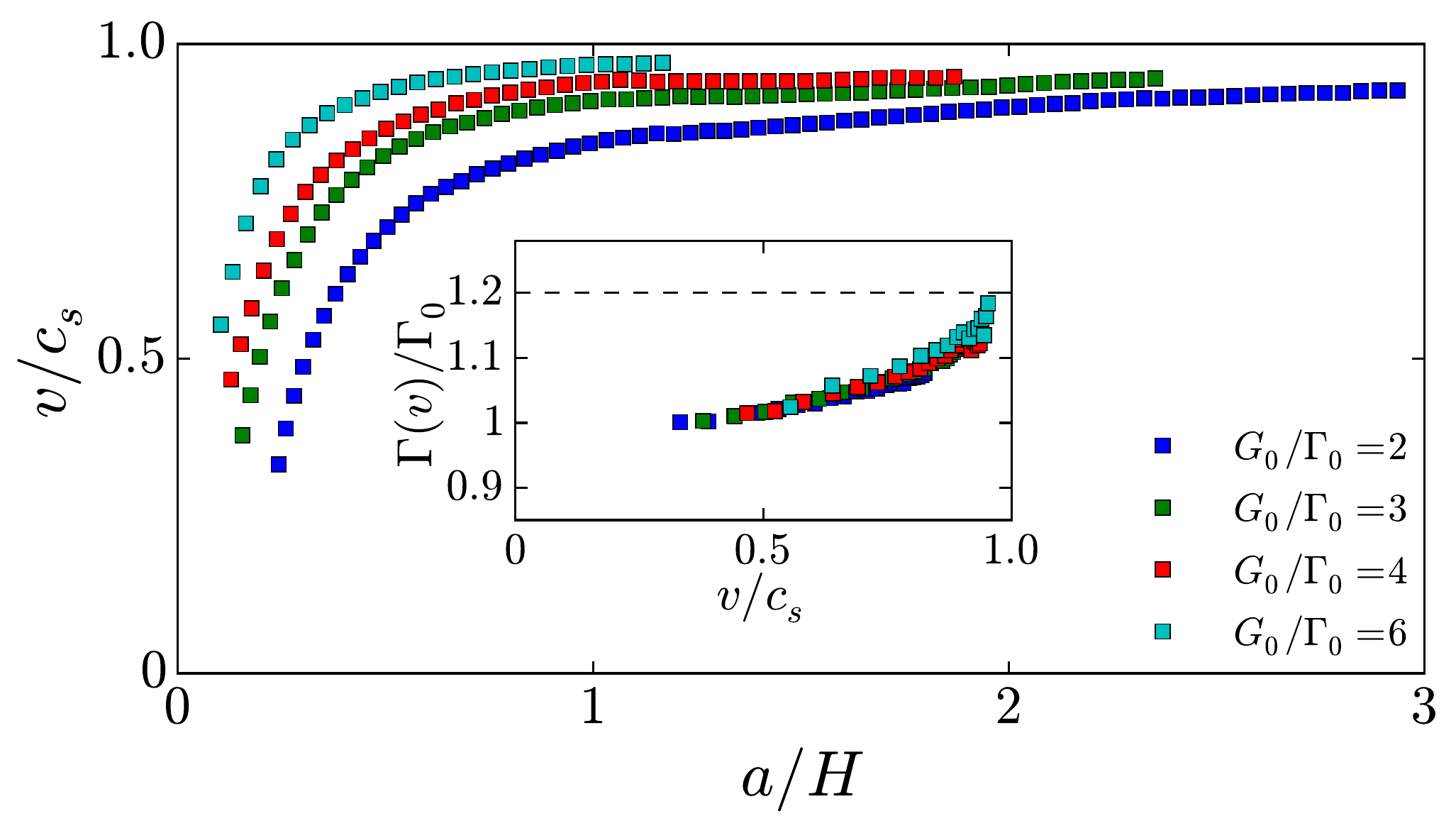}
\caption{
{\bf Accelerating crack simulations and the equation of motion for straight cracks.}
Results of accelerating crack simulations in the geometry of Fig. S3.
Plots of crack velocity $v$ (normalized by $c_s$) versus crack length $a$ (normalized by $H$) for different loads $G_0/\Gamma_0$ (see legend). The plots correspond to straight cracks prior to the onset of instabilities. Simulation parameters are $\ell_0/\xi=1.45$, $H/\xi\!=\!300$, $W/\xi\!=\!900$, $\beta\!=\!0.28$ and $\Delta\!=\!0.21\xi$. The fracture energy $\Gamma(v)\!=\!G$ (normalized by $\Gamma_0$) versus crack velocity $v$, obtained from $J$-integral calculations of the energy release rate $G$ (cf.~Eqs.~\eqref{J-integral}-\eqref{J-integral2}) (inset). All data, corresponding to different loading levels, collapse onto a unique fracture energy curve $\Gamma(v)$. The ratio $\Gamma(v)/\Gamma_0$ increases from 1 to 1.2 (dashed line) when $v$ varies from 0 to $v_c$.}
\label{figs4}
\end{center}
\end{figure}

In Fig.~S4, the crack velocity $v$ (normalized by $c_s$) is plotted versus the total crack length $a$ (normalized by $H$). Note that in the strip geometry of Fig.~S3, we chose to monitor the crack dynamics by measuring the total crack length $a$, which is the distance between two tips of a centered crack, while in the treadmill simulation geometry of Fig. S1, which was used to generate the results reported in Figs. 2, 3 and 4 of the main text, we monitored the propagation distance $d$ of a single crack tip. Fig.~S4 shows that different loading levels generate different crack velocity-length relations  $v(a)$ and consequently different crack acceleration histories. The crack velocity-length curves $v(a)$ are truncated at unprecedentedly high velocities before the onset of instabilities. The equation $G\!=\!\Gamma(v)$ predicts that even though $v(a)$ is different for each loading level, the energy release rate $G$ is a unique function of the instantaneous crack velocity $v$. To test this prediction, we calculated the energy release rate $G$ using the $J$-integral, cf.~Eqs.~\eqref{J-integral}-\eqref{J-integral2}, for the different $v(a)$ curves. 
The results presented in the inset of Fig.~S4 demonstrate that all curves collapse on a single curve defining the dynamic fracture energy $\Gamma(v)$ (normalized here by $\Gamma_0$), in complete agreement with the relation $G\!=\!\Gamma(v)$ for straight cracks.
The results are also consistent with the theoretical expectation that this relation, which simply accounts for energy balance near the crack tip, should remain valid even in the presence of elastic nonlinearity as long as $G$ is calculated through the $J$-integral evaluated in a non-dissipative region.

{\bf Limiting speed of cracks ---}
The onset velocity of oscillations $v_c\!\approx\!0.92\,c_s$ in Figs.~2 and 3a of the main text is in remarkably good quantitative agreement with experiments \cite{Livne.07} and is slightly less than the Rayleigh wave speed $c_R=0.933\,c_s$, which is the theoretical limiting speed of cracks predicted by LEFM for a linear elastic material. For the simulations reported in Fig. 3b and 4b of the main text, the crack velocity can slightly exceed the Rayleigh wave speed. This result is in accord with theoretical studies showing that the limiting velocity of surface acoustic waves approaches $c_s$ in the limit of large background strain in a pre-stretched incompressible neo-Hookean material \cite{destrade2007interface, dowaikh1990surface}, and with a theoretical analysis of near-tip stress fields showing that $c_s$ is the limiting speed of cracks in the same limit \cite{Livne.10}. Consistent with those theoretical predictions, we find that the limiting speed of cracks can slightly exceed $c_R$ in simulations when the background strain $\varepsilon_{yy}$ in the strip exceeds a few percent. 

{\bf Numerical regularization of the strain energy density  ---} During propagation of quasi-static model I cracks, all regions of the material are under tension and hence the out-of-plane stretch ratio $[{\det}({\bm F})]^{-1}$ appearing in Eq.~\eqref{eq:e_strain} is less than unity everywhere in space. However, under dynamic conditions, this ratio can become much larger than unity in regions behind the crack tip where the material becomes transiently highly compressed by nonlinear elastic waves. To prevent numerical issues associated with $[{\det}({\bm F})]^{-1}$ becoming too large, we modified the strain energy density to have the form
\begin{eqnarray}
e_{\mbox{\scriptsize{strain}}}=\frac{\mu}{2}\left(F_{ij}F_{ij}+\frac{1}{J^2}-3\right), \label{eq:e_strain_modified}
\end{eqnarray}
where
\begin{eqnarray}
\frac{1}{J^{2}} =
\begin{cases}
	{1}/\left[\det({\bm F})\right]^2, & ~~~~\text{if}~~ [\det({\bm F})]^{-1}<J_{\rm min}^{-1} \\
	8/\left[\det({\bm F})+J_{\rm min}\right]^{2}-1/J_{\rm min}^{2} , & ~~~~\text{otherwise}
\end{cases}
\label{eq:numerical}
\end{eqnarray}
where $J_{\rm min}$ is a numerical cutoff parameter. Eq.~\eqref{eq:numerical} regularizes the divergence of $e_{\mbox{\scriptsize{strain}}}$ in the $\det({\bm F})\rightarrow 0$ limit and ensures that $J$ and its first derivative are continuous at $\det({\bm F})\!=\!J_{\rm min}$. We used $J_{\rm min}\!=\!0.2$ in all simulations reported in this study. This cutoff, which is primarily introduced to ensure numerical stability, does not affect the crack dynamics that is predominantly controlled by tensile regions of the material where the out-of-plane stretch ratio $[\det({\bm F})]^{-1}<1$.
By repeating simulations with smaller values of $J_{\rm min}$ (e.g. $J_{\rm min}=0.05, 0.1$), we verified that the results are independent of the choice of this cutoff as long as $J_{\rm min}$ is chosen small compared to unity. 

\bibliography{fracture_Nature}